# Thickness gradient promotes the performance of Si-based anode material for lithium-ion battery

Zhenbin Guo[1,2], Haimin Yao[1,2,*]

[1]The Hong Kong Polytechnic University Shenzhen Research Institute, Shenzhen 518057, China

[2]Department of Mechanical Engineering, The Hong Kong Polytechnic University, Hung Hom, Kowloon, China

*Corresponding author at: The Hong Kong Polytechnic University Shenzhen Research Institute, Shenzhen 518057, China.
E-mail address: mmhyao@polyu.edu.hk (H. Yao).

## Abstract

The large volume change of the silicon (Si) during lithiation and delithiation process has long been a problem impeding its application as one of the most promising anode materials for LIBs. In this paper, we proposed a conceptually new idea to address this problem simply by varying the thickness of the electrode material film. The resulting thickness-gradient electrode exhibits considerable enhancement in the electrochemical performances including capacity, capacity retention, Coulombic efficiency, and rate capability in comparison to the traditional counterparts with uniform thickness. Such enhancement in the electrochemical performance can be attributed to the lessening of the stress concentration on the interface between the electrode film and the current collector upon the volume change of Si taking place in the lithiation and delithiation process. To make the best use of this strategy, optimal design of the gradient thickness is proposed based on the theory of stress homogenization, followed by the experimental verification. The results of this paper provide a facile, cost-effective, and scalable way for enhancing the performance of Si-based anodes for LIBs. This strategy can be further extended to the other anode materials suffering from the similar lithiation-induced volume change problem.

# 1. Introduction

Among diverse devices for energy storage, lithium-ion battery (LIB) stands out for its comparatively high capacity, better rate capability, and longer lifespan [1]. Nevertheless, recently with the ever-increasing demand for high-performance power supply in industries and personal electronics, the limitation of the capacity of the prevalent carbon-based LIBs starts to emerge [2]. Developing next-generation LIB with higher capacity and cyclability is believed to provide a mighty thrust to the development of technology and the global economy. A straightforward measure to improve the capacity of LIB is to replace the carbon with other electrode materials with a higher capacity [3-5]. However, a problem commonly existing in the anode materials with high capacity is the large volume change (LVC) during the lithiation and delithiation process, which will cause the pulverization of the electrode material film as well as detachment from the current collector [6-8]. As one the most representative examples, Silicon (Si) possesses theoretical specific capacity as high as 4200 mAh $g^{-1}$, which is one order of magnitude higher than that of the graphite [9, 10]. However, the lifespan of the Si-based LIBs is much shorter than that of the carbon-based counterparts. This can be attributed to the LVC (300-400%) of the Si during the lithiation and delithiation processes [11, 12]. To tackle the LVC problem of Si, a bunch of research efforts has been made [13-24], including the application of constraining coating on Si materials [25-27], structurization of the Si nanomaterials such as 0-dimensional nanoparticles [28, 29], 1-dimensional nanowires [30-32] and 2-dimensional thin films [33, 34]. Despite the success of these approaches in alleviating the LVC problem of Si [35-38], the sophisticated chemical processes and expensive fabrication facilities involved greatly limit the transfer of the laboratory techniques to the existing LIB industry. To find a facile, cost-effective, and scalable solution to the LVC problem of Si, we cast our attention to the interface between the electrode film and current collector.

From a mechanical point of view, the interface between the electrode film and the underlying current collector is the Achilles heel of the bilayer structure, because the strain misfit across the interface, which may be induced by the volume change of the electrode film during lithiation and delithiation, will lead to significant stress and strain concentration and singularity [39-42]. Interfacial delamination tends to occur after sufficient cycles of charging and discharging, resulting in the degradation of the electrochemical performance. Suppressing the interfacial stress and strain concentration is believed beneficial for solving of the LVC problem of Si. Inspired by the success of functionally graded material (FGM) in alleviating the interfacial stress concentration [43-45], we had purposely altered the concentration of Si nanoparticles in the thickness direction of the electrode film. The resulting graded electrode exhibits elevated electrochemical performance in comparison to the traditional ones [46]. The gain of such performance enhancement involves no additional materials and chemical processes. This success implies the great promise of gradient thickness as another strategy for homogenizing interfacial stress discovered recently [47].

To testify the feasibility of the gradient thickness in addressing the LVC problem of Si-based anode, in this paper we deliberately change the thickness of a circular Si-based anode film, which is traditionally uniform, to be descending along the radial direction, as shown schematically in Fig. 1a. In view of the practical difficulty in fabricating such continuously variable thickness, an approximate substitute is proposed by stacking multiple thinner sublayers with different and descending diameters, resulting in electrode with stepwise gradient (Fig. 1b). The remainder of this paper is structured as follows. Firstly, the detailed manufacturing process and electrochemical characterization are introduced. Then, the obtained results are discussed to show the effectiveness of the thickness gradient in promoting the electrochemical performance of LIBs. To make the best use of this strategy, the optimal design of the gradient thickness is explored in both theory and experiment. Finally, the paper is concluded by comparing the current strategy with similarities discovered earlier.

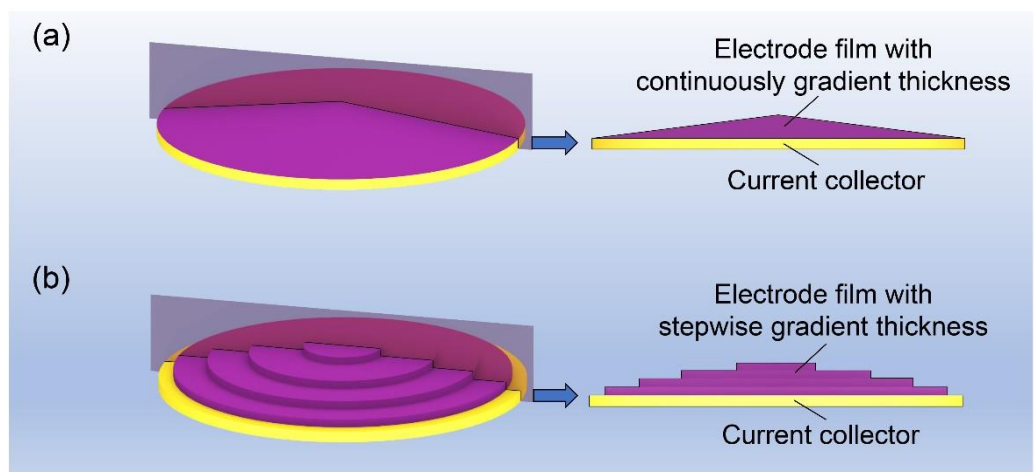

**Fig. 1** Schematics of electrodes with (a) continuously gradient thickness, and (b) stepwise gradient thickness.

## 2. Experimental procedure

2.1 *Preparation of Si-based uniform electrodes for LIB through spray painting*

Traditionally, the electrode material of LIB is cast onto the current collector (Cu foil) by using a film applicator. The thickness of the resulting film of the electrode material is normally uniform. To prepare electrode film with gradient thickness, spray painting technique is applied with an apertured disk used as the mask. The process is elaborated as follows.

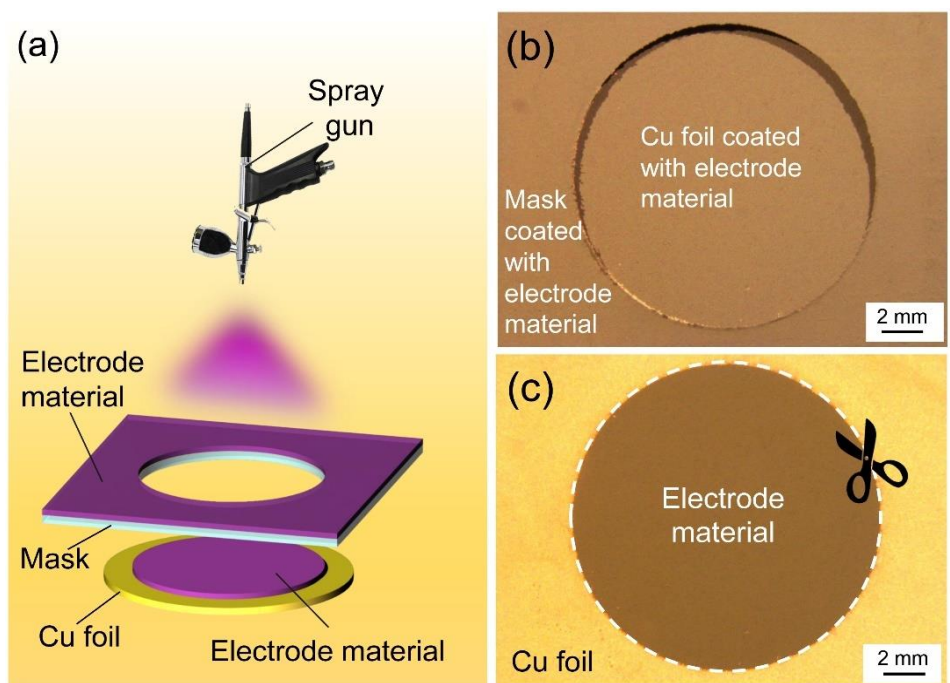

**Fig. 1** (a) Schematic illustration of the spraying-based method for fabricating electrodes for LIB; (b) Optical microscope image (top view) of a masked Cu foil after spraying and drying; (c) Optical microscope image (top view) of a Cu foil deposited with a circular film of Si-based electrode material.

Firstly, the Si-based electrode material is prepared by mixing Si NPs, carbon black (conductive agent), and PVDF (binder) with solvent N-Methyl-2-pyrrolidone (NMP) in a mass ratio of 3:1:1:30 using pestle and mortar. Then, the as-prepared slurry is filled into the container of the spray gun and spraying is carried out towards the copper (Cu) foil with an apertured mask applied amid to control the size of the region that can receive the spray on the Cu foil (Fig. 2a). The thickness of the electrode film is controlled by tuning the spraying time, while other affecting parameters such as the slurry concentration, air pressure, and distance between the spray gun and Cu foil, are kept fixed. After the spraying, the Cu foil together with the deposited electrode film and the mask is dried at 80 °C for 3 h followed by 10 h at 120 °C in a vacuum oven. The area of the dried electrode film on the Cu foil exhibits similar size to that of the aperture on the mask (Fig. 2b). After removing the mask and punchcutting the electrode film together with the underlying Cu foil (Fig. 2c), a disk-like Si-based electrode for LIB is obtained. The subsequent measurement indicates that the thickness of the prepared electrode film (dried) grows with the spraying time in a nonlinear manner (Fig. 3a). Initially, the growth rate is 0.65 μm/s. As the spraying proceeds, the growth rate decreases and finally saturates at 0.05 μm/s.

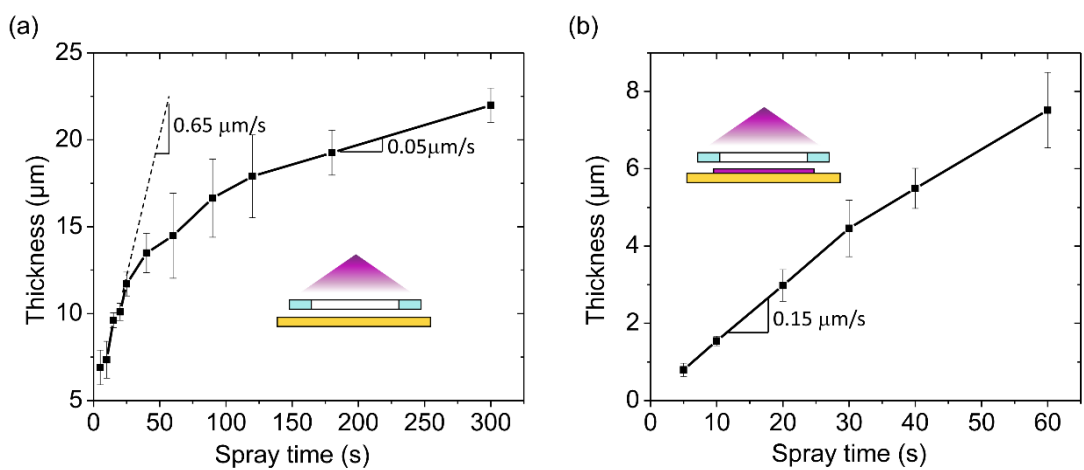

**Fig. 2** Variation of the thickness of the deposited film with the spraying time for (a) spraying directly on Cu foil, and (b) spraying on a pre-deposited and dried electrode film.

2.2 *Preparation of Hanoi-tower-like electrodes through multiple spraying*

The method of spraying developed above allows us to prepare electrode film with gradient thickness by multiple steps of spraying in combination with the application of masks with descending aperture sizes. The resulting electrode film exhibits thickness with a stepwise gradient depending on the aperture size and the number of the masks applied and the spraying time in each step (Fig. 4a). Due to the resemblance to the Hanoi-tower in morphology, the electrode with such stepwise thickness is called Hanoi-tower-like (HTL) electrode hereafter. Fig. 4b shows the evolution of the electrode film (dried) after 1, 2, and 3 steps of spraying. To prepare the HTL electrodes in a controllable way, the thickness growth in each spraying step should be characterized quantitively in advance. Surprisingly, if the spraying is conducted on a pre-deposited and dried electrode film rather than on a Cu foil, the thickness grows linearly with the spraying time at a growth rate of 0.15 μm/s (Fig. 3b). This is different from the growth rate of the electrode film directly deposited on the Cu foil, which might be attributed to the re-wetting and deformation of the pre-deposited film caused by the subsequent spraying.

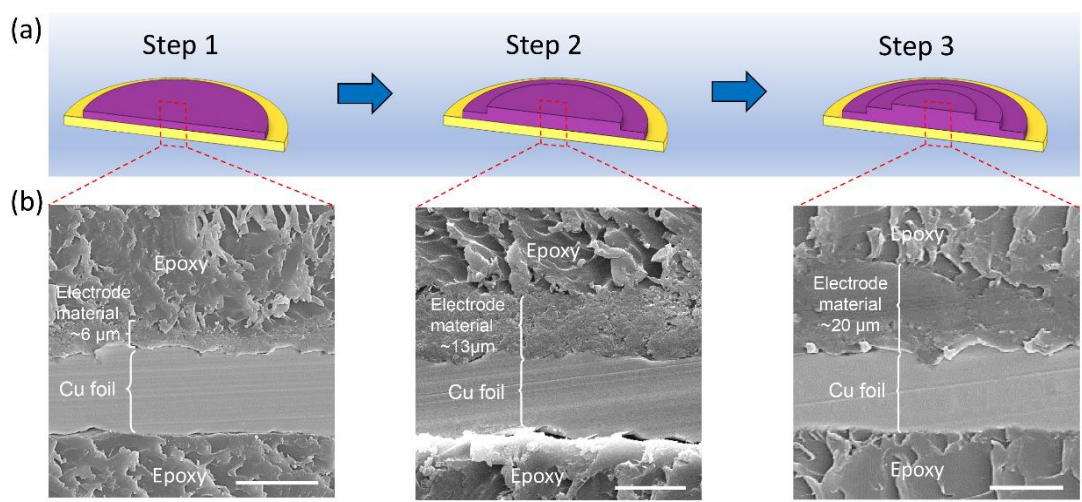

**Fig. 4** Cross-sectional SEM images of the electrode showing the thickness of electrode film after each step of spraying. Scale bar: 15 μm.

2.3 *Electrochemical characterization*

The electrochemical performance of the as-prepared electrode is characterized using CR2032 coin-type half-cells with Lithium foil being employed as the counter electrode, Celgard 2400 as the separator, 1 M LiPF6 in ethylene carbonate and diethyl carbonate (EC:DEC=1:1) with a 5 vol.% fluoroethylene carbonate (FEC) and 1 vol.% vinylene carbonate (VC) additive as the electrolyte. Galvanostatic discharge/charge test is carried out with a LAND CT-2001A at the potential ranging from 0.01 to 1.2 V vs. Li$^+$/Li. All electrochemical cycling measurements are carried out at room temperature. All the specific capacities and current density are assessed based on the weight of Si applied.

## 3. Results and discussion

Three-layer (3L) and seven-layer (7L) HTL electrodes were prepared and characterized by using the multiple spraying method introduced above (see Table S1 and Table S2 for the detailed fabrication parameters). For comparison, electrodes with the same constituents and mass loading but uniform thickness were prepared, including the uniform electrodes made by casting and spraying methods, respectively. The average specific charge capacity of the half-cells of each type of electrodes is shown in Fig. 5a-b. Comparing two types of uniform electrodes made by different methods, the electrochemical performance shows certain dependence on the manufacturing method. In particular, an electrode made by casting performs better than that made by spraying especially in the initial 20 cycles. The average specific charge capacity (*N*=5) of the uniform electrodes by casting is 568 mAh g$^{-1}$ after 10 cycles and will dry out after 30 cycles, while the average specific charge capacity (*N*=5) of the uniform electrode made by spraying is 198 mAh g$^{-1}$ after 10 cycles and will dry out after 20 cycles. In comparison to the electrodes with uniform thickness, the 3L-HTL electrode exhibits a much higher specific capacity. The average specific charge capacity ($N$=3) of the 3L-HTL electrodes is 1408 mAh g$^{-1}$ after 10 cycles, 906 mAh g$^{-1}$ after 50 cycles, and 684 mAh g$^{-1}$ after 100 cycles, respectively. The benefit of gradient thickness can be further extracted if we increase the layer number from 3 to 7. In particular, the average specific charge capacity ($N$=3) of the 7L-HTL electrodes is 2574 mAh g$^{-1}$ at the first cycle and 976 mAh g$^{-1}$ after 100 cycles.

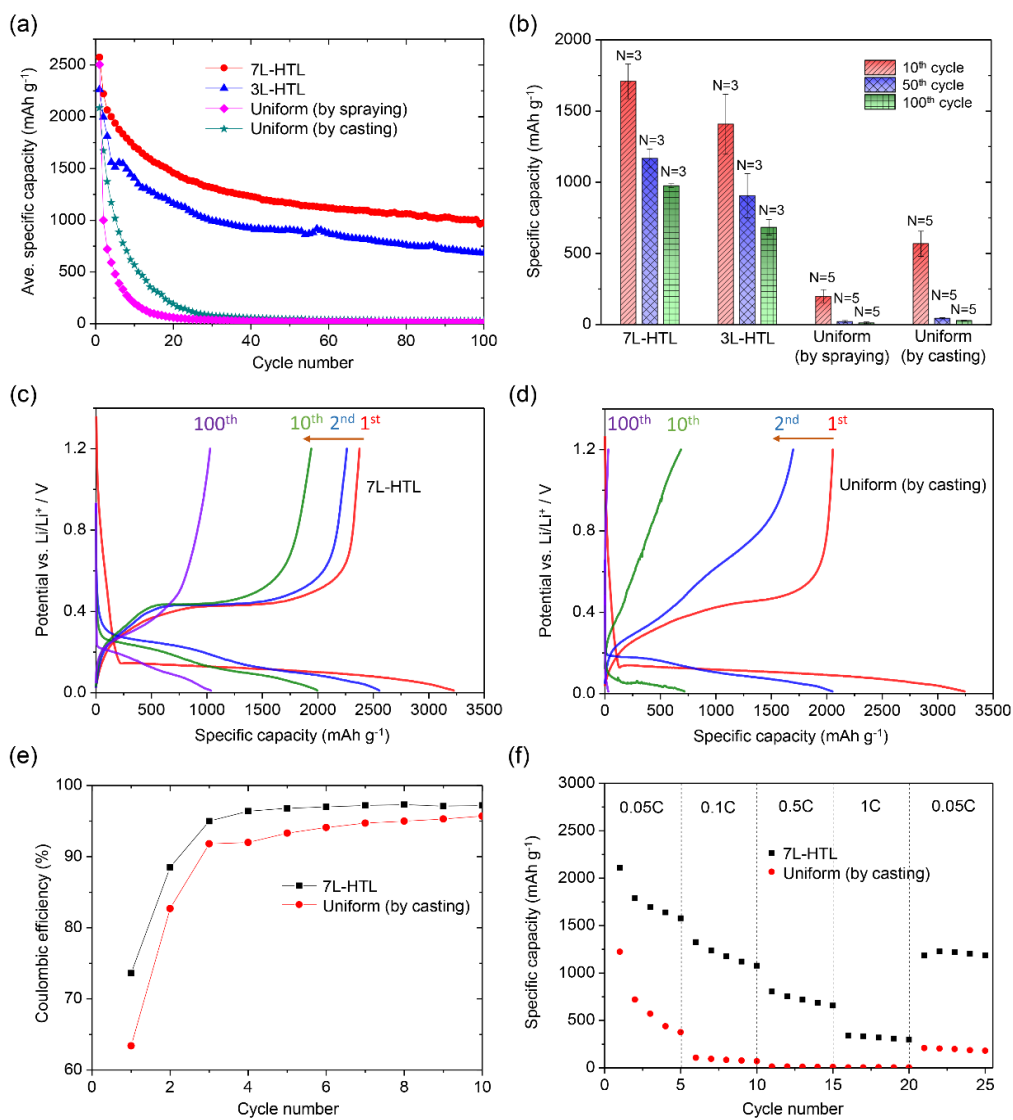

**Fig. 5** (a) The average specific charge capacity of the 7L-HTL electrodes ($N$=3), 3L-HTL electrodes ($N$=3), uniform electrodes made by spraying ($N$=5) and uniform electrodes made by casting (*N*=5) during cycling test at the voltage between 0.01 V and 1.2 V vs. Li/Li$^+$ and a rate of 0.05 C; (b) Comparison of the specific capacity at the selected cycles. Here, *N* is the number of tested half-cells; The 1$^{st}$, 2$^{nd}$, 10$^{th}$ and 100$^{th}$ galvanostatic discharge/charge profiles of (c) a 7L-HTL electrode and (d) a uniform electrode made by casting. (e) Coulombic efficiency of a 7L-HTL electrode in comparison to that of a uniform electrode made by casting. (f) The rate capability of a 7L-HTL electrode in comparison to that of a uniform electrode made by casting at a rate varying from 0.05 C to 1 C.

Fig. 5c shows the galvanostatic discharge/charge profiles of the 7L-HTL electrode at the voltage range of 0.01-1.20 V and a rate of 210 mA g$^{-1}$. When Li-ions get inserted into the electrode with gradient thickness for the first time (discharging or lithiation process), the voltage drops drastically to 0.14 V, followed by a steady plateau and then a progressive decrease to 0.01 V. During the ensuing charging (delithiation) process, the voltage initially increases quickly, followed by a relatively steady plateau ranging from 0.22 V to 0.48 V. Similar plateaus can be observed even after 100 cycles, implying a stable voltage window for charging and discharging. In contrast, for the uniform electrode made by casting, the voltage plateau is only observed in the first a few cycles and disappears after 10 cycles (Fig. 5d). Moreover, the initial discharge and charge capacity of the 7L-HTL electrode is 3225 and 2377 mAh g$^{-1}$, respectively. The initial Coulombic efficiency is 74% which is higher than that of the uniform electrode made by casting (~63%) (Fig. 5e). This implies that the electrode film with gradient thickness can effectively enhance the utilization efficiency of Si in the electrode.

The rate capability of the 7L-HTL electrode, which reflects the degradation of performance of the LIB at a higher charge/discharge rate, is compared with that of the conventional uniform electrode in Fig. 5f. At the same discharge/charge rate varying from 0.05 C to 1 C (1 C = 4200 mA g$^{-1}$), the 7L-HTL electrode always delivers a higher specific capacity than the uniform one does. When the rate returns to 0.05 C, the specific capacity of the 7L-HTL electrode is 1184 mAh g$^{-1}$, which greatly overpasses that of the uniform one (208 mAh g$^{-1}$). These results indicate that the electrode film with gradient thickness can significantly improve the rate capability of Si-based LIB.

The success of the HTL electrode in enhancing the cycling performance of LIB motivates us to explore the optimal design of the gradient thickness that can bring us the maximum profit of this strategy. Since the mechanism accounting for such success is the lessening of the stress concentration and singularity by the gradient thickness, maximum profit is expected if one can achieve a completely uniform shear stress field on the interface when strain misfit is present. This question actually has been systematically discussed in our previous work for different configurations of bi-material systems [47]. Specifically, for a circular electrode film, the optimal thickness profile of the film is analytically given by

$$t_e(r) = \frac{(1-v_e^2)E_c t_c}{(1-v_c^2)E_e} \left\{ \left[ \frac{(v_e+2)(1-v_c^2) + \left[\frac{3E_c t_c \varepsilon_{mis}}{R\tau_{exp}} - (v_c+2)(1-v_c)\right](v_e+1)}{(v_e+2)(1-v_c^2) \cdot \frac{r}{R} + \left[\frac{3E_c t_c \varepsilon_{mis}}{R\tau_{exp}} - (v_c+2)(1-v_c)\right](v_e+1)} \right]^{\frac{3}{2+v_e}} - 1 \right\} \quad (1)$$

where $r$ is the distance from the film center; $E_e$ and $v_e$ are Young's modulus and Poisson's ratio of the electrode film, respectively; $E_c$ and $v_c$ are Young's modulus and Poisson's ratio of the current collector, respectively; $t_c$ and $R$ are the thickness and radius of the current collector, respectively; $\varepsilon_{mis}$ is the strain misfit between the electrode film and current collector (Cu foil) caused by the intrinsic volume change of Si during the process of lithiation and delithiation; $\tau_{exp}$ is the expected homogeneous interfacial shear stress at strain misfit of $\varepsilon_{mis}$. In physics, $\tau_{exp}$ is capped by the shear strength of the interface. For given $E_e$, $\nu_e$, $E_c$, $\nu_c$, $t_c$, and $R$, Eq.(1) indicates that the gradient thickness $t_e(r)$ is correlated to the ratio of $\tau_{exp}/\varepsilon_{mis}$, which represents the growth rate of the expected homogeneous shear stress on the interface with the increase of strain misfit, and therefore can be deemed as the shearing stiffness of the interface corresponding to that gradient thickness.

Taking $E_e = 19.7$ GPa [48], $\nu_e = 0.3$ (assumed), $E_c = 128$ GPa, $\nu_c = 0.33$ [49], $t_c = 10$ μm, $R = 7.5$ mm in Eq. (1), the cross-sectional profile of the optimal thickness $t_e(r)$ is plotted in Fig. 6a for different shearing stiffnesses of $\tau_{exp}/\varepsilon_{mis}$. It can be seen that the cross-sectional profile of the optimal thickness exhibits a tent-like shape. The higher the tent, the greater the resulting interfacial shearing stiffness $\tau_{exp}/\varepsilon_{mis}$.

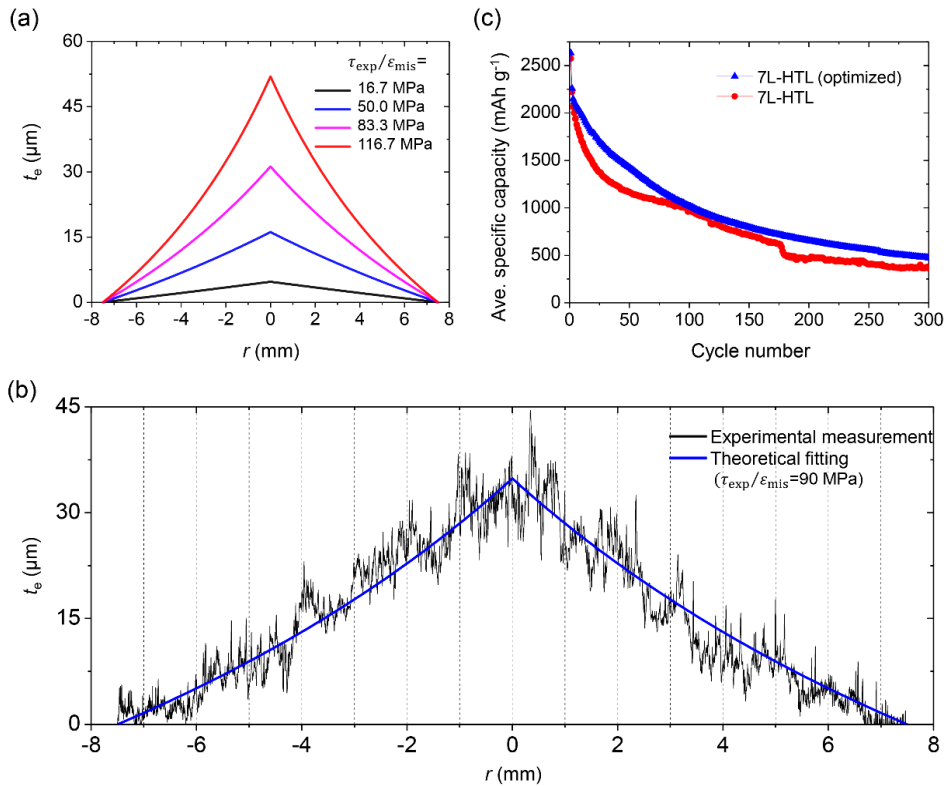

**Fig. 6** (a) The optimal profiles of electrode films corresponding to different shearing stiffnesses of the interface ($\tau_{exp}/\varepsilon_{mis}$). Here, the base diameter of the circular electrode is taken as 15 mm. (b) The measured topography of an optimized 7L-HTL electrode. (c) The average specific charge capacity ($N$=3) of the optimized 7L-HTL electrodes in comparison to that of the unoptimized ones.

Following the guideline by the theoretical solution to the optimal profile of the gradient thickness, we prepared an optimized 7L-HTL (see Table S3 for the detailed fabrication parameters). Topographical characterization with a surface profiler (Dektak XT) along the radial direction indicates that the manufactured profile is quite close to the theoretical solution to the optimal profile corresponding to $\tau_{exp}/\varepsilon_{mis} = 90$ MPa (Fig. 6b). Subsequent electrochemical characterization on the optimized 7L-HTL electrode exhibits a better cycling performance compared to that of the unoptimized ones (Fig. 6c). In particular, the average specific charge capacity ($N$=3) of the optimized 7L-HTL electrodes is 2633 mAh g$^{-1}$ at the first cycle, 1990 mAh g$^{-1}$ after 10 cycles, 1019 mAh g$^{-1}$ after 100 cycles and 472 mAh g$^{-1}$ after 300 cycles, respectively. Such enhancement in the electrochemical performance also outruns the enhancement brought by the gradient in Si concentration as discovered in our previous study [46] (Fig. 7a).

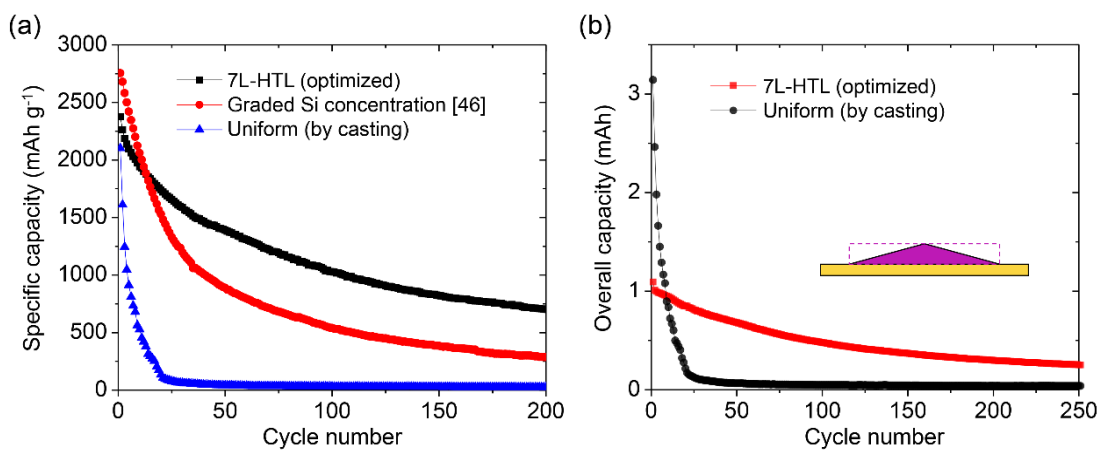

**Fig. 7** (a) Comparison of the cycling performance of an electrode with gradient thickness (optimized 7-HTL) and an electrode with gradient in Si composition developed in our previous study [46] ; (b) Comparison of the cycling performance of a 7L-HTL electrode with a uniform electrode (as depicted by the broken line in the inset) with the same height (~15 μm).

For a practical battery cell, what is more important is its overall capacity, which reflects the total amount of energy that can be stored, rather than the specific capacity of the active material. This is because the specific capacity of the active material (*e.g*., Si) tends to decrease as the mass loading increases. The preceding discussion has demonstrated the advantage of gradient thickness over the uniform thickness in the aspect of specific capacity. It is practically worthy of making a comparison between the gradient and uniform electrodes from the perspective of overall capacity. For this purpose, we compare the overall capacities of an optimized 7L-HTL electrode (see Table S4 for the detailed fabrication parameters) and a uniform control with the same film height (Fig. 7b). It can be seen that the overall capacity of the uniform electrode drops quickly with the increasing cycle number. After about 9 cycles, the overall capacity of the uniform electrode drops below that of the 7L-HTL electrode even though more active material (Si) is contained in the uniform electrode than in the gradient one. Therefore, the gradient electrode achieves a higher overall capacity with the use of less active material. This feature implies the great promise and value of the gradient-thickness electrode for the cost control in the LIB industry.

## 4. Conclusion

In this paper, we tackled the LVC problem of the Si-based electrodes simply by introducing the thickness gradient into the electrode film. The resulting gradient electrodes with descending thickness exhibit much better cycling performance in comparison to the traditional uniform ones as well as the ones with graded Si concentration. The optimal design of the gradient thickness is discovered, allowing us to make the best use of the strategy of gradient thickness. In comparison to the other strategies for addressing the LVC problem of Si, this strategy is facile to implement and cost-effective and involves no additional chemical processes and materials. The scaling up of this method in the existing battery industry is relatively easy. Admittedly, the electrode films that we fabricated above with the spraying-based method achieved only a stepwise gradient in thickness. For electrode films with continuous gradient, which may bring additional enhancement to the electrochemical performance, sophisticated technique such as additive manufacturing should be adopted. Our results in this paper provide a conceptually new idea for addressing the LVC problem of Si and is believed to promote the application of Si-based anode materials in the next-generation LIBs.

**Acknowledgments**

This work is supported by the National Natural Science Foundation of China (11772283).

**Declaration of Competing Interest**

None.

**Supplementary Information**

Table S1. Fabrication parameters of the 3L-HTL electrodes

| Sublayer Number | Aperture diameter (mm) | Spraying time* (s) | Drying time (h) | Sublayer thickness (dried) (μm) |
|---|---|---|---|---|
| 1st | 15 | 10 | 3 | 6 |
| 2nd | 10 | 60 | 3 | 7 |
| 3rd | 5 | 60 | 3 | 7 |

*The distance between the spraying nozzle and Cu foil is 20 cm.

Table S2. Fabrication parameters of the 7L-HTL electrodes

| Sublayer Number | Aperture diameter (mm) | Spraying time* (s) | Drying time (h) | Sublayer thickness (dried) (μm) |
|---|---|---|---|---|
| 1st | 14 | 5 | 3 | 4 |
| 2nd | 12 | 25 | 3 | 4 |
| 3rd | 10 | 25 | 3 | 4 |
| 4th | 8 | 25 | 3 | 4 |
| 5th | 6 | 25 | 3 | 4 |
| 6th | 4 | 25 | 3 | 4 |
| 7th | 2 | 25 | 3 | 4 |

*The distance between the spraying nozzle and Cu foil is 20 cm.

Table S3. Fabrication parameters of the optimized 7L-HTL electrodes

| Sublayer Number | Aperture diameter (mm) | Spraying time* (s) | Drying time (h) | Sublayer thickness (dried) (μm) |
|---|---|---|---|---|
| 1st | 14 | 3 | 3 | 2 |
| 2nd | 12 | 13 | 3 | 2 |
| 3rd | 10 | 20 | 3 | 3 |
| 4th | 8 | 25 | 3 | 4 |
| 5th | 6 | 35 | 3 | 5 |
| 6th | 4 | 45 | 3 | 6 |
| 7th | 2 | 55 | 3 | 7 |

*The distance between the spraying nozzle and Cu foil is 20 cm.

**Table S4**. Fabrication parameters of the optimized 7L-HTL electrodes with the height of ~15 μm

| Sublayer Number | Aperture diameter (mm) | Spraying time* (s) | Drying time (h) | Sublayer thickness (dried) (μm) |
|---|---|---|---|---|
| 1st | 14 | 3 | 3 | 2 |
| 2nd | 12 | 13 | 3 | 2 |
| 3rd | 10 | 13 | 3 | 2 |
| 4th | 8 | 13 | 3 | 2 |
| 5th | 6 | 13 | 3 | 2 |
| 6th | 4 | 13 | 3 | 2 |
| 7th | 2 | 20 | 3 | 3 |

*The distance between the spraying nozzle and Cu foil is 20 cm.